# $B$-type Quadratic Planar Hall Effect as a Probe of Altermagnetic Order

Yiwei Zhao[1], Junwei Liu[2, *], Jian Zhou[3, *], Haowei Xu[1, *]

[1] *Department of Physics, City University of Hong Kong, Kowloon 999077, Hong Kong, China*

[2] *Department of Physics, The Hong Kong University of Science and Technology, Hong Kong, China*

[3] *Center for Alloy Innovation and Design, State Key Laboratory for Mechanical Behavior of Materials, Xi'an Jiaotong University, Xi'an 710049, China*

## Abstract

Conventional transport and optical responses, such as the Hall effect, can only distinguish material phases with substantially different symmetries. However, many quantum materials, such as candidate altermagnetic materials, host competing phases with similar structures, making it difficult to differentiate them using conventional transport or optical measurements. Here, guided by symmetry principles, we introduce a $B$-type quadratic planar Hall effect ($B^2$-PHE) as a sensitive probe of magnetic structures. $B^2$-PHE features a planar Hall current that scales quadratically with the applied magnetic field. This response originates from the interplay among external magnetic fields, intrinsic magnetic order, and quantum geometric properties of the electronic bands. We investigate $B^2$-PHE in two altermagnetic candidates, $KV_2Se_2O$ and $RuO_2$, whose ground-state magnetic orders are under intense debate. Remarkably, $B^2$-PHE emerges in their altermagnetic phases but is symmetry-forbidden in their antiferromagnetic (nonmagnetic) phases. We also show that $B^2$-PHE should be detectable under a moderate magnetic field on the order of 1 Tesla. These findings establish $B^2$-PHE as an experimentally convenient and potentially unambiguous probe of altermagnetic order, which is otherwise difficult to realize.

***Introduction.*** Transport and optical responses are among the most convenient and powerful probes of quantum materials, enabling the characterization of electronic structures [1-6], magnetic orders [7-9], and other symmetry-driven quantum phenomena [10-17]. These responses are most effective at distinguishing material phases with *substantially different symmetries*. For example, the anomalous Hall effect identifies phases that break time-reversal symmetry [18-21], whereas second-harmonic generation (SHG) is sensitive to broken spatial inversion symmetry or changes in rotational symmetry [22-25]. However, many quantum materials exhibit competing phases that share *highly similar symmetries*, rendering them largely indistinguishable by conventional transport or optical techniques.

A prominent example is altermagnetic (AM) materials, which have recently ignited intense interest. They combine ferromagnetic-like properties, such as efficient spin transport, with antiferromagnetic-like advantages, including ultrafast terahertz dynamics and the absence of stray fields [20, 26-41]. Despite this promise, the experimental identification of AM order in many materials remains highly controversial because these candidate materials host competing magnetic orders (Table S1 in the Supplementary Materials, SM). For instance, $RuO_2$ was long regarded as a prototypical altermagnet [42, 43], but subsequent transport and scattering experiments indicate that it may be nonmagnetic (NM) [44-46]. Similar discrepancies have also emerged in layered compounds such as $KV_2Se_2O$, which belongs to the $(Rb, K)_{1-\delta}V_2(Se, Te)_2O$ family [34]. Spectroscopy and neutron diffraction experiments have reached conflicting conclusions regarding their AM versus antiferromagnetic (AFM) orders [41, 47].

A straightforward symmetry analysis reveals that conventional transport and optical responses, such as the Hall effect and SHG, cannot readily distinguish the competing magnetic phases of the candidate altermagnets discussed above (Table I and Table S2 in SM). For example, the Hall conductivity tensor of $RuO_2$ ($KV_2Se_2O$) has the same symmetry-allowed form in both its AM and NM (AFM) phases, while SHG is forbidden in both phases. These limitations underscore the need for additional transport and/or optical phenomena that are intrinsically sensitive to the underlying magnetic orders, so that AM and competing magnetic states can be unambiguously distinguished.

In this regard, we propose a $B$-type quadratic planar Hall effect ($B^2$-PHE), guided by symmetry principles. $B^2$-PHE is a nonlinear transport phenomenon characterized by a planar Hall current that scales quadratically with the applied magnetic field, that is, $j_a \propto B_c B_d$, where $a$, $c$, and $d$ denote Cartesian indices. This should be distinguished from common nonlinear transport or optical effects, which scale nonlinearly with the electric field. Intuitively, because both $B_c$ and $B_d$ couple directly to intrinsic magnetic moments, $B^2$-PHE is expected to be highly sensitive to the underlying magnetic orders. This expectation is confirmed by our symmetry analysis and first-principles calculations. Remarkably, $B^2$-PHE is symmetry-allowed in the AM phases of both $RuO_2$ and $KV_2Se_2O$, reaching experimentally detectable magnitudes under a moderate magnetic field of $\sim 1$ Tesla. In contrast, it is strictly forbidden by symmetry in the corresponding AFM or NM phases. These results establish $B^2$-PHE as a convenient, sensitive, and potentially unambiguous transport probe for distinguishing AM order from competing magnetic phases.

In the following, we first perform a comprehensive symmetry analysis of the transport and optical responses of candidate altermagnets in their AM and AFM (NM) phases, highlighting the unique capability of the $B^2$-PHE to distinguish between the competing magnetic orders. Then, we elucidate the microscopic mechanism of $B^2$-PHE, revealing its intimate correlation with magnetic order and quantum geometric properties of the electronic structure. This microscopic mechanism also enables an efficient first-principles framework to calculate $B^2$-PHE in real materials. We apply this framework to representative AM candidates, $KV_2Se_2O$ and $RuO_2$, and discuss the experimental feasibility.

**Table I.** Symmetry analysis of various transport and optical effects in AM and AFM (NM) phases of candidate altermagnets, $RuO_2$ and $KV_2Se_2O$. In the Jahn symbol, $V$ denotes a vector, $a$ and $e$ denote time-reversal odd and axial transformation characters. "Same" indicates that the response tensors have the same symmetry-allowed form, so their symmetry-allowed forms do not distinguish the relevant phases. We included only magnetic structures with zero net magnetic moments. Similar analyses are provided in SM Table S2 for $RuO_2$ and $KV_2Se_2O$ with magnetic moments in the $x$-$y$ plane, as well as for other AM candidates, such as MnTe.

| Jahn symbol | | $[V^2]$ | $a\{V^2\}$ | $eV\{V^2\}$ | $V[V^2]$ | $aV[V^2]$ | $V[V^3]$ | $\boldsymbol{a\{V^2\}[V^2]}$ |
|---|---|---|---|---|---|---|---|---|
| Representative effects | | Dielectric response | Anomalous Hall effect | Conventional Hall effect | Second-harmonic generation | Nonlinear Hall effect | Third-harmonic generation | **$B^2$-PHE (this work)** |
| $RuO_2$ | AM along $z$ ($4'/mm'm$) | Same (Table S3) | Forbidden | Same (Table S4) | Forbidden | Forbidden | Same (Table S5) | **Allowed** (Table S6) |
| | NM ($4/mmm$) | | | | | | | **Forbidden** |
| $KV_2Se_2O$ | AM along $z$ ($4'/mm'm$) | Same (Table S3) | Forbidden | Same (Table S4) | Forbidden | Forbidden | Same (Table S5) | **Allowed** (Table S6) |
| | AFM along $z$ ($4/mmm.1'$) | | | | | | | **Forbidden** |

***Symmetry Analysis.*** We first perform a symmetry analysis of various transport and optical responses, taking $RuO_2$ as an example. The AM phase of $RuO_2$ belongs to the magnetic point group $4'/mm'm$ when the Néel vector $L$ is along the $z$ axis ($L \parallel \hat{z}$), whereas the NM phase belongs to $4/mmm$. Symmetry analysis shows that these two phases exhibit remarkably similar transport and optical responses (Table I). For instance, both the anomalous Hall effect and SHG are forbidden, while the conventional Hall effect is characterized by response tensors of the same form in both phases. Consequently, these conventional responses cannot be used to distinguish the two phases. Note that $RuO_2$ might have other magnetic configurations as well, such as ferromagnetic, ferrimagnetic, or AM states with $L \parallel \hat{x}$ or $\hat{y}$. These configurations may host finite magnetic moments or possess lower magnetic symmetry, so they can be distinguished more readily (Table S2 in SM). Hence, we only focus on AM ($L \parallel \hat{z}$) and NM phases here.

The discussions above motivate us to propose the $B^2$-PHE, guided by symmetry principles. Notably, the AM ($L \parallel \hat{z}$) and NM phases $RuO_2$ are allowed and forbidden by symmetry for responses characterized by the Jahn symbol $aV^4$, that is, time-reversal-odd third-order responses (see caption of Table I). Although this symmetry class encompasses multiple transport or optical phenomena, we focus on a Hall current that is quadratic in the magnetic field and linear in the electric field, because it is experimentally straightforward to realize. Specifically, one has

$$j_a = \sigma^{(2)}_{abcd} E_b B_c B_d, \tag{1}$$

where $j$, $E$, and $B$ denote charge current density, electric field, and magnetic field, respectively, while $a, b, c,$ and $d$ are Cartesian indices. Note that $\sigma^{(2)}_{abcd}$ is antisymmetric under $a \leftrightarrow b$ and symmetric under $c \leftrightarrow d$, so $B^2$-PHE has a Jahn symbol of $a\{V^2\}[V^2]$.

Symmetry analysis shows that in the AM ($L \parallel \hat{z}$) phase of $RuO_2$, the only nonzero components of the conductivity $\sigma^{(2)}_{abcd}$ are (Table S6 in SM)

$$\sigma^{(2)}_{xyxx} = -\sigma^{(2)}_{yxxx} = \sigma^{(2)}_{yxyy} = -\sigma^{(2)}_{xyyy} = c_1 \tag{2}$$

$$\sigma^{(2)}_{xz[yz]} = -\sigma^{(2)}_{zx[yz]} = \sigma^{(2)}_{yz[xz]} = -\sigma^{(2)}_{zy[xz]} = c_2 \tag{3}$$

where $c_1$ and $c_2$ are two independent coefficients, and the notation $[cd]$ denotes that $c$ and $d$ can be exchanged (symmetric). In contrast, in the NM phase of $RuO_2$, all $\sigma^{(2)}_{abcd}$ components are symmetry-forbidden and thus vanish identically. Similarly, $\sigma^{(2)}_{abcd}$ is symmetry-allowed in the AM phase but forbidden in the AFM phases of $KV_2Se_2O$. Such a striking contrast implies that the $B^2$-PHE is a powerful probe for identifying the AM ($L \parallel \hat{z}$) phase in candidate altermagnets.

Note that there can be a Hall current that scales linearly with magnetic field as well (i.e., conventional Hall effect), which in practice should be isolated from the quadratic contribution to better distinguish the competing phases. It is thus advantageous to consider the $c_1$ components in Eq. (2), for which the current, electric field, and magnetic field all lie in the same $x$-$y$ plane. The linear Hall current in this coplanar configuration corresponds to the planar Hall effect (PHE) [48-53]. Importantly, the linear PHE is symmetry-forbidden in both the AM ($L \parallel \hat{z}$) and AFM (NM) phases of $RuO_2$ and $KV_2Se_2O$, leaving the quadratic contribution as the leading-order Hall response in this geometry. In the following, we will thus focus on this response and refer to it as the $B$-type quadratic planar Hall effect ($B^2$-PHE).

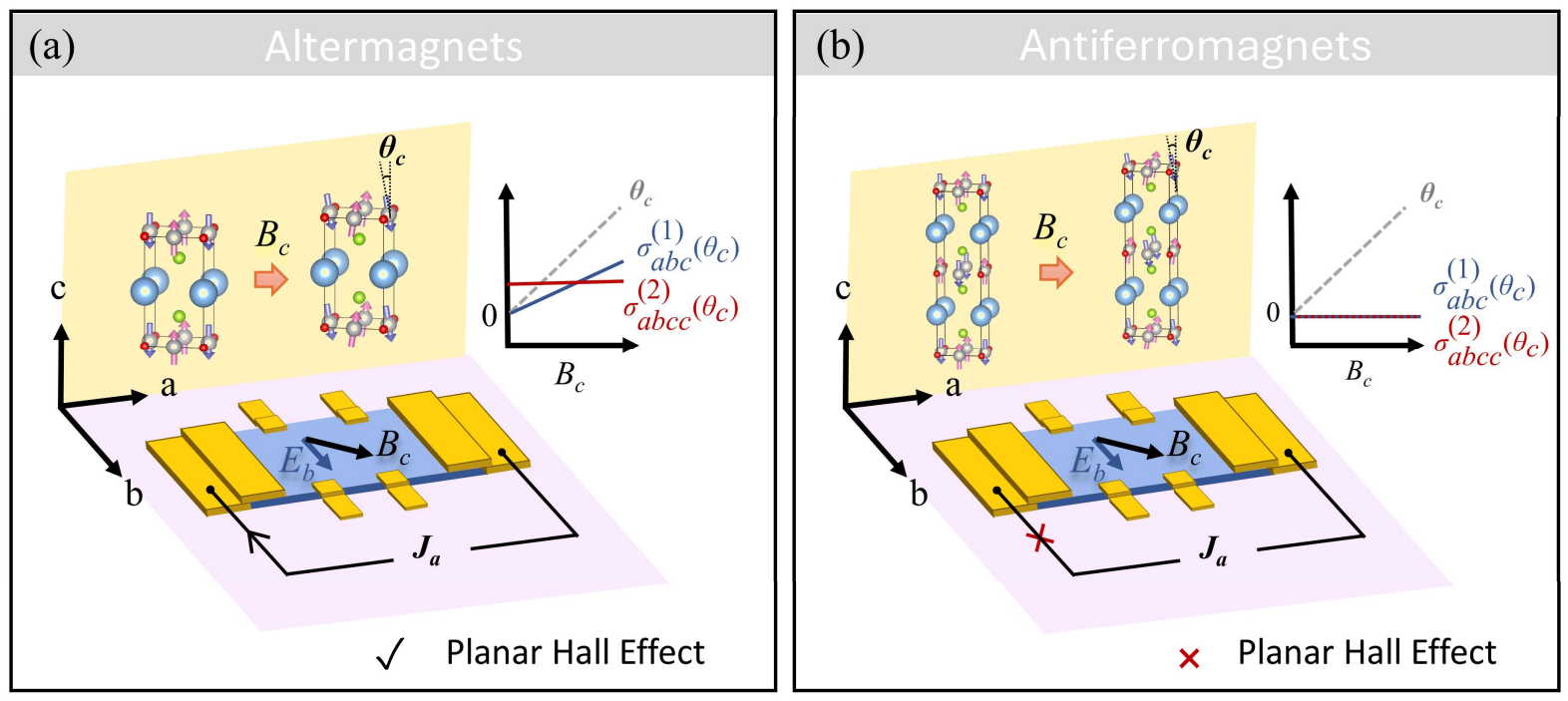

**FIG. 1**. Schematic illustration of the $B$-type quadratic planar Hall effect ($B^2$-PHE), which is allowed by symmetry in (a) the AM phase but forbidden in (b) the AFM phase of $KV_2Se_2O$. $B^2$-PHE is also allowed in the AM phase but forbidden in the NM phase of $RuO_2$.

**Microscopic mechanism of $B^2$-PHE.** Next, we briefly discuss the microscopic mechanism underlying the $B^2$-PHE. In magnetic materials, a primary influence of an external transverse magnetic field is to cant the magnetic moments $M_0$. In AM $KV_2Se_2O$, for example, the magnetic moments align along the $z$ direction in equilibrium. When a magnetic field is applied along $c = x$ or $y$, the moments cant towards the $c$ direction by an angle of $\theta_c \approx \frac{M_c}{M_0} = \chi B_c$ (Figure 1), where the canting coefficient $\chi$ can be related to the magnetic susceptibility. This linear relationship between $\theta_c$ and $B_c$ is valid in the weak-field regime. The spin canting lowers the magnetic point group symmetry of AM $KV_2Se_2O$ from $4'/mm'm$ to $2'/m'$, thereby activating the linear PHE. The linear PHE generates a Hall current $j_a = \sigma^{(1)}_{abc} E_b B_c$, where $a$ and $b$ are orthogonal by the definition of the Hall effect, and $c$ lies in the plane spanned by $a$ and $b$ ($x$-$y$ plane here). As we will elaborate later, the linear PHE conductivity in canted AM $KV_2Se_2O$ satisfies $\sigma^{(1)}_{abc}(\theta_c) \propto \theta_c \propto B_c$. Comparing $\sigma^{(2)}_{abcc}$ of un-canted structures with $\sigma^{(1)}_{abc}(\theta_c)$ of canted structures, one has

$$\sigma^{(2)}_{abcc} = \frac{\sigma^{(1)}_{abc}(\theta_c)}{B_c} = \frac{\chi \sigma^{(1)}_{abc}(\theta_c)}{\theta_c} \tag{4}$$

Phenomenologically, one of the $B_c$ terms in Eq. (1) cants the magnetic moments by a small angle $\theta_c$, while the other $B_c$ term induces the linear PHE current in the canted structure with a conductivity of $\sigma_{abc}^{(1)}(\theta_c)$. This relation provides a convenient route for the first-principles calculations of $\sigma_{abcc}^{(2)}$. Specifically, we first perform density functional theory (DFT) calculations with the magnetic moments constrained to a series of canting angles $\theta_c$ and obtain the corresponding electronic structures. We then compute the PHE conductivity $\sigma_{abc}^{(1)}(\theta_c)$ using a formalism that has been established previously [50, 52, 53] (see below). Finally, $\sigma_{abcc}^{(2)}$ is extracted from a linear fit of $\sigma_{abc}^{(1)}(\theta_c)$ as a function of $\theta_c$, according to Eq. (4). The canting coefficient $\chi$ can be determined from the total energy dependence on $\theta_c$, as described in the SM. Note that due to the relationship in Eq. (2), we only need to consider one component, namely, $\sigma_{xyxx}^{(2)} = \frac{\sigma_{xyx}^{(1)}(\theta_x)}{B_x}$.

The linear PHE conductivity $\sigma_{abc}^{(1)}(\theta_c)$ can be evaluated using the extended semiclassical theory. In essence, external fields induce corrections to the electronic band properties, resulting in a charge current [50, 52]. Within this formalism, the conductivity is given by [50]

$$\sigma_{xyc}^{(1,\alpha)} = \frac{e^2}{\hbar} \sum_n \int \frac{d^3k}{(2\pi)^3} f_n' \left[ v_x^{nn} \mathcal{A}_{y,c}^{\alpha,n} - v_y^{nn} \mathcal{A}_{x,c}^{\alpha,n} - \Omega_{xy}^n m_c^{\alpha,nn} \right]. \tag{5}$$

Here $\int \frac{d^3k}{(2\pi)^3}$ denotes the integration over the first Brillouin zone, $n$ is the band index, and $f_n'$ is the derivative of Fermi-Dirac distribution with respect to electron band energy. We define the magnetic-field-corrected Berry connection $\mathcal{A}_{a,c}^{\alpha,n} = -2\hbar \mathrm{Im} \sum_{m \neq n} \frac{v_a^{nm} m_c^{\alpha,mn}}{(\varepsilon_n - \varepsilon_m)^2}$ with $\hbar$ the reduced Planck constant and the Berry curvature $\Omega_{xy}^n = -2\mathrm{Im} \sum_{m \neq n} \frac{v_x^{nm} v_y^{mn}}{(\varepsilon_n - \varepsilon_m)^2}$. Meanwhile, $\varepsilon_n$ is the band energy, while $v^{mn}$ and $m^{\alpha,mn}$ are the matrix elements of the velocity and magnetic moment operators, respectively. We use superscript $\alpha = \mathrm{sp}$ or $\mathrm{orb}$ to denote the spin or orbital contributions, and the total response is $\sigma_{xyc}^{(1)} = \sigma_{xyc}^{(1,\,\mathrm{sp})} + \sigma_{xyc}^{(1,\,\mathrm{orb})}$. Specifically, the spin magnetic moment is $m_c^{\mathrm{sp},mn} = -g\mu_B s_c^{mn}$, where $g = 2$ is the electron $g$-factor, $\mu_B$ is the Bohr magneton, while $s_c$ is the $c$-component Pauli operator. The orbital magnetic moment can be obtained from the modern theory of orbital magnetization [52], $m_c^{\mathrm{orb},mn} = \frac{e}{2\hbar} \left[ \sum_{l \neq n,m} \boldsymbol{v}^{ml} \times \boldsymbol{A}^{ln} + (\boldsymbol{v}^{nn} + \boldsymbol{v}^{mm}) \times \boldsymbol{A}^{mn} \right]_c$, where $\boldsymbol{A}^{ln}$ is the interband Berry connection. One can see that the PHE is closely related to quantum geometric properties [50, 52, 53].

Our symmetry analysis above shows that one should have $\sigma_{xyc}^{(1)}(\theta_c) \propto \theta_c \propto B_c$ for both $RuO_2$ and $KV_2Se_2O$, which is also verified by our first-principles calculations, as shown later. Here, we further provide an intuitive microscopic explanation for this relationship, using Eq. (5). We first note that the magnetic moment operator $m_c^{mn}(\theta_c)$ is a function of $\theta_c$, and there are three relationships: (1) the canted part of the permanent magnetic moment is $M_c = \sum_n m_c^{nn}(\theta_c) \propto \theta_c$, suggesting a linear relationship between $m_c^{nn}(\theta_c)$ and $\theta_c$; (2) the linear PHE with $\theta_c = 0$ is $\sigma_{xyc}^{(1)}(0) = \sum_{k,mn} \mathcal{F}\{m_c^{mn}(0)\} = 0$ for both $RuO_2$ and $KV_2Se_2O$. Here $\mathcal{F}\{m_c^{mn}(\theta_c)\}$ is a linear function of $m_c^{mn}(\theta_c)$ [see Eq. (5)]. This indicates that (3) $\sigma_{xyc}^{(1)}(\theta_c \neq 0) = \sum_{k,mn} \mathcal{F}\{m_c^{mn}(\theta_c)\}$ comes solely from the difference $\delta m_c^{mn} \equiv m_c^{mn}(\theta_c) - m_c^{mn}(0)$. These three relationships imply $\sigma_{xyc}^{(1)}(\theta_c) \propto \theta_c$.

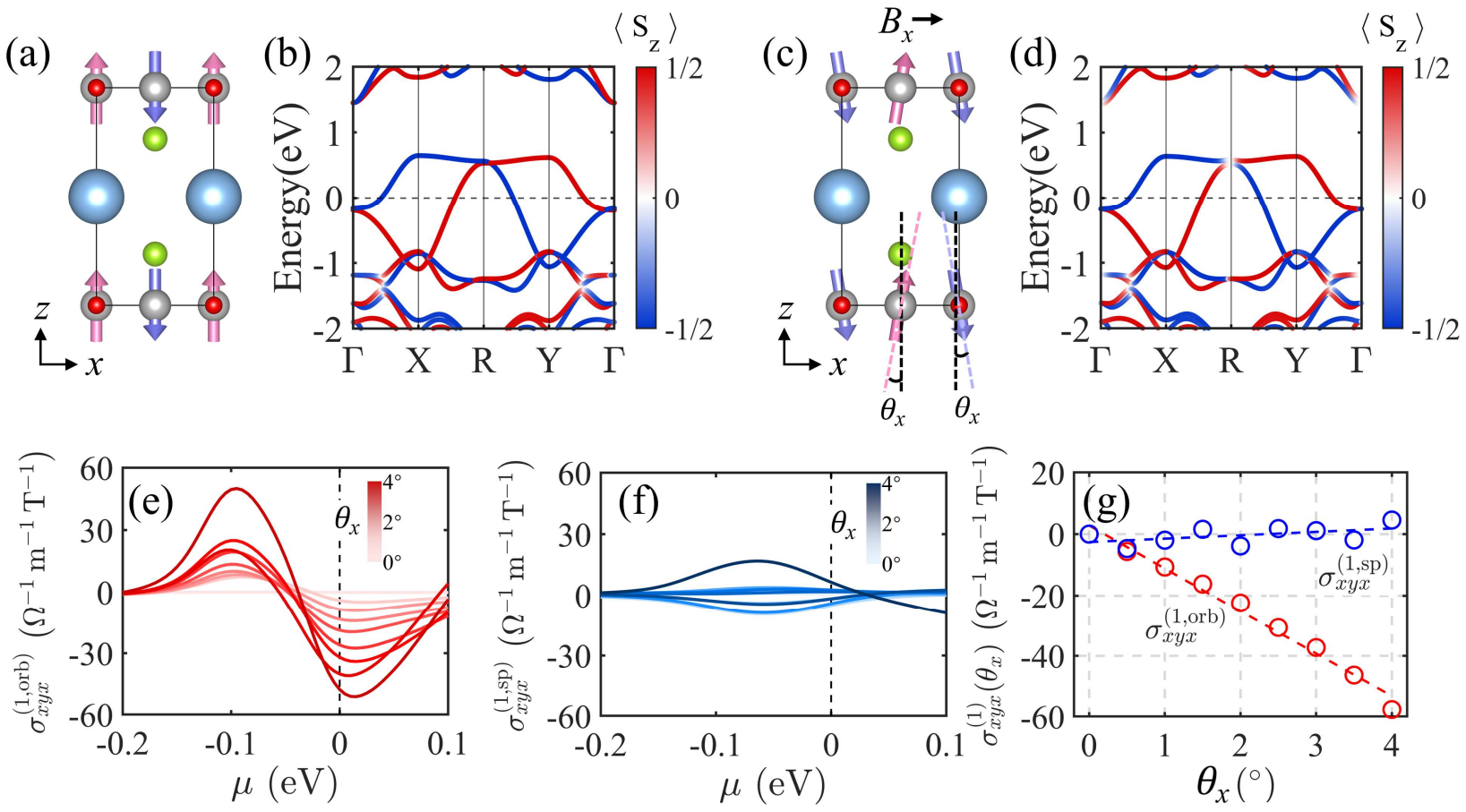

**FIG. 2.** (a) Crystal and magnetic structure of bulk $KV_2Se_2O$ in the AM ($L \parallel \hat{z}$) phase. Pink and blue arrows indicate opposite magnetic moments on V sites. V, Se, K and O atoms are depicted by gray, green, blue and red balls. (b) Electronic band structure with canting angle $\theta_x = 0°$, with color map indicating the band-resolved spin polarization $\langle S_z \rangle$. (c) Canting of magnetic moments under an in-plane $B_x$ with a canting angle of $\theta_x$. (d) Same as (b), but for $\theta_x = 4°$. (e) Orbital and (f) spin contributions to the planar Hall conductivity, $\sigma_{xyx}^{(1,\mathrm{orb})}$ and $\sigma_{xyx}^{(1,\mathrm{sp})}$, as functions of chemical potential $\mu$ for different canting angles $\theta_x$. The vertical dashed line marks the Fermi level. (g) Linear planar Hall conductivity with $\mu = 0$ as a function of $\theta_x$. Note that both $\sigma_{xyx}^{(1,\mathrm{orb})}$ and $\sigma_{xyx}^{(1,\mathrm{sp})}$ are zero for the pristine AM phase with $\theta_x = 0°$.

***$B^2$*-PHE in AM and AFM $KV_2Se_2O$**. In the following, we will demonstrate how $B^2$-PHE can be used to distinguish AM from competing magnetic orders. We first consider bulk $KV_2Se_2O$ in the AM ($L \parallel \hat{z}$) phase, which hosts compensated collinear magnetic moments along the $z$ direction on V atoms [Fig. 2(a)]. It exhibits a pronounced, momentum-dependent non-relativistic spin splitting in the electronic band structure [Fig. 2(b)], the tell-tale feature of AM ordering. This equilibrium AM ($L \parallel \hat{z}$) state belongs to the high-symmetry magnetic point group $4'/mm'm$, for which the intrinsic linear PHE is forbidden by symmetry. This is confirmed by our first-principles calculations, as shown in Figs. 2(e-g) when $\theta_x = 0°$.

Then, we cant the magnetic moments toward the $x$-axis by an angle of $\theta_x$, resulting in a net magnetic moment per V atom, $M_x \approx M_0\theta_x$, where $M_0 = 2\mu_B$ is the total magnetic moment on each V atom [Fig. 2(c)]. This reduces the magnetic point group to $2'/m'$. Such a symmetry reduction removes the constraints that forbid the linear PHE, thereby allowing for a finite $\sigma_{xyx}^{(1)}$ for $\theta_x \neq 0$. The canted magnetic structure with a typical $\theta_x = 4°$ exhibits a visible band splitting near the high-symmetry points in the Brillouin zone, as shown in Fig. 2(d). Figures 2(e) and 2(f) show the orbital and spin contributions to $\sigma_{xyx}^{(1)}(\theta_x)$ as functions of chemical potential $\mu$, for various $\theta_x$ up to $4°$. As expected, $\sigma_{xyx}^{(1,\mathrm{orb})}(\theta_x)$ increase almost linearly with $\theta_x$, in agreement with the microscopic mechanism discussed above. It reaches a sizable value of around $57\ \Omega^{-1}\ \mathrm{m}^{-1}\ \mathrm{T}^{-1}$ at $\mu = 0$ and $\theta_x = 4°$. Meanwhile, the spin contribution $\sigma_{xyx}^{(1,\mathrm{sp})}$ has a much smaller magnitude at $\mu = 0$. By fitting $\sigma_{xyx}^{(1)}$ as a function of $\theta_x$ [Eq. (4)], we obtain the $B^2$-PHE conductivity as $\sigma_{xyxx}^{(2)} \sim 1\ \Omega^{-1}\ \mathrm{m}^{-1}\ \mathrm{T}^{-2}$. This corresponds to a Hall conductivity of approximately $1\ \Omega^{-1}\ \mathrm{m}^{-1}$ under an in-plane magnetic field of $B_x \sim$ 1 T, indicating that the predicted $B^2$-PHE should be experimentally detectable.

To further confirm the symmetry selectivity of $B^2$-PHE, we also consider the recently debated AFM phase of $KV_2Se_2O$, which consists of a doubled unit cell with antiparallel magnetic moments in adjacent V layers (SM Fig. S1). The magnetic point group of AFM $KV_2Se_2O$ is $4/mmm.1'$, which does not allow $B^2$-PHE. Correspondingly, the canted AFM state ($\theta_x \neq 0$) belongs to the magnetic point group $m'm'm$, forbidding the emergence of linear PHE. This is verified by our first-principles calculations, showing that $\sigma_{xyx}^{(1)}(\theta_x)$ remains zero for any $\theta_x \neq 0$. These results suggest that the $B^2$-PHE can be a convenient and potentially unambiguous probe of the AM and AFM ordering in $KV_2Se_2O$.

Interestingly, the proposed $B^2$-PHE probe of AM order naturally extends to the two-dimensional limit. Bilayer $KV_2Se_2O$ belongs to the magnetic point groups $4'/mm'm$ and $4/mmm.1'$ in the AM and AFM phases, respectively. Therefore, $B^2$-PHE is allowed in the AM but forbidden in the AFM phase, which is also verified by our calculations (Fig. S2).

***B*²-PHE in AM and NM $RuO_2$.** We next apply our framework to bulk $RuO_2$ in the AM ($L \parallel \hat{z}$) phase, which features compensated magnetic moments along the $z$ direction on Ru atoms [Figure 3(a)]. Similar to AM $KV_2Se_2O$, the electronic band structure exhibits a clear momentum-dependent spin splitting [Fig. 3(c)].

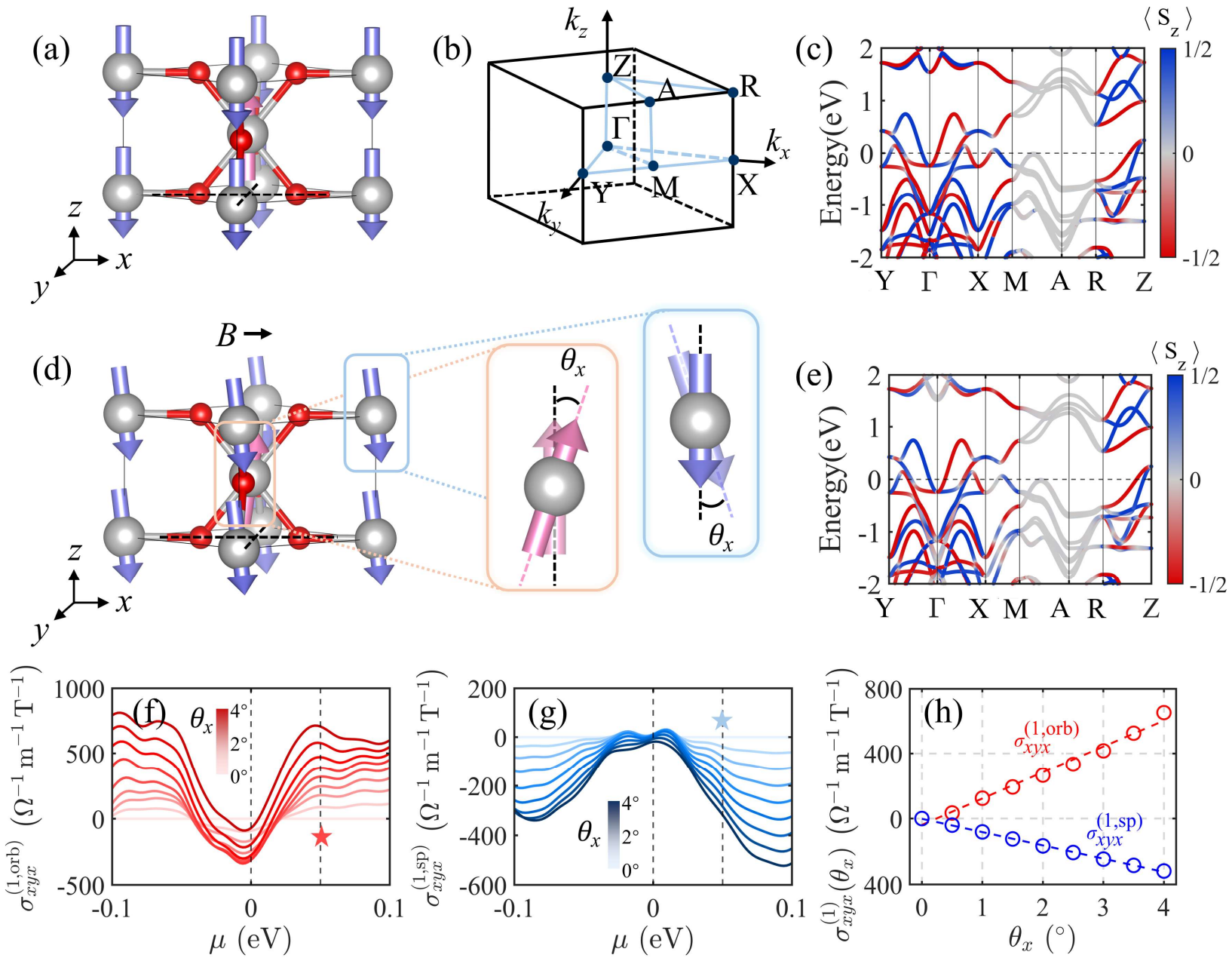

**FIG. 3**. (a) Crystal and magnetic structure of bulk $RuO_2$ in the AM ($L \parallel \hat{z}$) phase. Pink and blue arrows indicate opposite magnetic moments on Ru sites. Ru and O atoms are depicted by gray and red balls. (b) First Brillouin zone of $RuO_2$ with high-symmetry points. (c) Electronic band structure of $RuO_2$ with $\theta_x = 0^\circ$, with the color map indicating the band-resolved spin polarization $\langle S_z \rangle$. (d) Canting of the magnetic moments under $B_x$ with canting angle denoted by $\theta_x$. (e) Same as (c), but for $\theta_x = 4^\circ$. (f) Orbital and (g) spin contributions to the planar Hall conductivity, $\sigma_{xyx}^{(1,\mathrm{orb})}$ and $\sigma_{xyx}^{(1,\mathrm{sp})}$, as functions of chemical potential $\mu$ for different canting angles $\theta_x$. The vertical dashed lines mark $\mu = 0$ eV and $\mu = 0.05$ eV. (h) Linear planar Hall conductivity at $\mu = 0.05$ eV as a function of $\theta_x$.

We then cant the magnetic moments towards the $x$-direction, characterized by a canting angle $\theta_x$. The band structure at $\theta_x = 4^\circ$ is shown in Fig. 3(e). The canting reduces the magnetic point group of $RuO_2$ from $4'/mm'm$ to $2'/m'$, thereby allowing for linear PHE. Figs. 3(f) and 3(g) show the orbital and spin contributions to the linear PHE conductivity $\sigma_{xyx}^{(1)}$ as functions of chemical potential $\mu$ for different canting angles $\theta_x$. As expected, both $\sigma_{xyx}^{(1,\mathrm{sp})}$ and $\sigma_{xyx}^{(1,\mathrm{orb})}$ increase linearly with $\theta_x$. The total response $\sigma_{xyx}^{(1)}$ can reach a few hundred $\Omega^{-1}\ \mathrm{m}^{-1}\ \mathrm{T}^{-1}$ at $\theta_x = 4^\circ$ for a wide range of $\mu$. This corresponds to a $B^2$-PHE conductivity of $\sigma_{xyxx}^{(2)} \sim 10\ \Omega^{-1}\ \mathrm{m}^{-1}\ \mathrm{T}^{-2}$, which is around one order of magnitude larger than that of $KV_2Se_2O$. We thus expect a more significant $B^2$-PHE feature in $RuO_2$ if it has an AM ($L \parallel \hat{z}$) order. In contrast, in the NM phase, the $B^2$-PHE remains zero due to symmetry constraints, as discussed before and shown in Fig. S3.

**Discussion.** In summary, we propose an unconventional $B^2$-PHE as a symmetry-sensitive probe of AM order in candidate altermagnets, including $RuO_2$ and $KV_2Se_2O$. This novel nonlinear transport mechanism is allowed in the AM phases but forbidden by symmetry in the corresponding competing AFM or NM phases. Using first-principles calculations, we find that the $B^2$-PHE conductivity in the AM phase can reach $1 \sim 10\,\Omega^{-1}\,\mathrm{m}^{-1}\,\mathrm{T}^{-2}$. This enables an efficient and potentially unambiguous experimental probe of the AM order under a moderate in-plane magnetic field, which has been challenging using other approaches. As a transport response, the $B^2$-PHE is suitable for metallic AM candidates. For semiconducting or insulating AM candidates, such as MnTe, whose magnetic structure is also under debate, one can introduce carriers or use optical responses as the probe. We will investigate these possibilities in future studies.

More broadly, our results demonstrate how fundamental symmetry principles can guide the on-demand design of novel transport and optical responses as selective probes of competing crystal, electronic, and magnetic phases in quantum materials. Since many quantum materials host competing phases with highly similar symmetries, we anticipate that this symmetry-guided approach will find broad applications beyond identifying AM orders.

**ACKNOWLEDGMENTS.** Y.Z. and H.X. acknowledge the support from the Hong Kong Research Grants Council under ECS Grant No. 21303526, National Natural Science Foundation of China under Grant No. 12604782, and the startup funding from City University of Hong Kong. J.Z. acknowledges the support from the National Natural Science Foundation of China under Grant No. 12374065. J. L. acknowledges the support from the Hong Kong Research Grants Council (CRS_HKUST603/25, C6046-24G, 16306722, 16304523 and 16311125).